\documentclass{iau} 
\usepackage{graphicx}
\usepackage{graphicx}
\usepackage{natbib}
\usepackage{color}
\newcommand{\araa}{ARA\&A}
\newcommand{\apj}{ApJ}
\newcommand{\apjl}{ApJ}
\newcommand{\aap}{A\&A}
\newcommand{\aapr}{A\&AR}
\newcommand{\jcap}{JCAP}
\newcommand{\mnras}{MNRAS}
\newcommand{\pasj}{PASJ}
\newcommand{\nat}{Nature}

\newcommand{\msun}{\,{\rm M_\odot}}

\newcommand{\beq}{\begin{equation}}
\newcommand{\eeq}{\end{equation}}

\title[S~319.~~Galaxies at High Redshift and Their Evolution over Cosmic Time] 
{The evolution of high-redshift massive black holes}

\author[Marta Volonteri, Melanie Habouzit, Fabio Pacucci, Michael Tremmel]   
{Marta Volonteri$^1$, Melanie Habouzit$^1$,  Fabio Pacucci$^2$ \and Michael Tremmel$^3$}

\affiliation{$^1$Institut dÕAstrophysique de Paris, Sorbonne Universit\`{e}s, UPMC Univ Paris 6 et CNRS, UMR 7095, 98 bis bd Arago, 75014 Paris, France \\ email: {\tt martav@iap.fr, habouzit@iap.fr} \\[\affilskip]
$^2$Scuola Normale Superiore, Piazza dei Cavalieri, 7  56126 Pisa, Italy \\email: {\tt fabio.pacucci@sns.it}\\[\affilskip]
$^3$Department of Astronomy, University of Washington, Seattle, WA \\email: {\tt mjt29@astro.washington.edu}}

\pubyear{2015}
\volume{xxx}  
\jname{Galaxies at High Redshift and Their Evolution over Cosmic Time}
\editors{A.C. Editor, B.D. Editor \& C.E. Editor, eds.}
\begin{document}

\maketitle

\begin{abstract}
Massive black holes (MBHs) are nowadays recognized as integral parts of galaxy evolution. Both the approximate proportionality between MBH and galaxy mass, and the expected importance of feedback from active MBHs in regulating star formation in their host galaxies point to a strong interplay between MBHs and galaxies. MBHs must form in the first galaxies and be fed by gas in these galaxies, with continuous or intermittent inflows that, at times, can be larger than the Eddington rate. Feedback from supernovae and from the MBHs themselves modulates the growth of the first MBHs. While current observational data only probe the most massive and luminous MBHs, the tip of the iceberg, we will soon be able to test theoretical models of MBH evolution on more ``normal" MBHs: the MBHs that are indeed relevant in building the population that we observe in local galaxies, including our own Milky Way.

\end{abstract}

\firstsection 
\section{Introduction}
Quasars have now been detected up to $z=7$ \citep{Mortlock2011}. The current sample, based on optical and near-infrared data, is characterized by high luminosity and large estimated black hole (BH) masses \citep[see Fig. 1 in][]{Wu2015}. Some fainter candidates have been proposed, based on X-ray observations \citep{2012A&A...537A..16F,2015A&A...578A..83G}, but they have not been confirmed yet \citep{2015MNRAS.448.3167W}. The MBHs powering these quasars have masses in excess of $10^8 \msun$, reaching $10^{10} \msun$. They are as massive as the largest MBHs today, but were in place when the Universe was less than one billion years old. The properties of quasars at $z>6$ require MBHs to form early on and grow rapidly. An initial MBH with mass 200 $\msun$ must grow continuously at the Eddington rate to reach $10^9 \msun$ by $z=7$.  A seed MBH with mass $10^5 \msun$ needs to accrete ``only" at half of the Eddington rate for the entire time, or at the Eddington rate for a half of their life. In contrast to such monsters, the MBH population extends down to small masses, though this range is hard to probe. The record for the smallest MBH currently belongs to the dwarf galaxy  RGG~118, which is thought to contain a black hole weighing only $\sim$ 50,000 $\msun$ \citep{2015ApJ...809L..14B}. There are also galaxies bereft of MBHs \citep[e.g., M33, NGC~205,][]{Gebhardt2001,Valluri2005}. Therefore, models of MBH formation, feeding and feedback must explain both the advent of the first luminous quasars,  why some galaxies host a MBH, and some others do not, and how different MBHs are fed, while providing at the same time the AGN feedback required to suppress star formation in massive galaxies. In this contribution we briefly review the formation, feeding and feedback of MBHs, with an emphasis on the high-redshift Universe \citep[see also][]{Volonteri2012Sci}.

\section{BH formation}

\begin{figure}[t]
\begin{center}
 \includegraphics[width=\columnwidth]{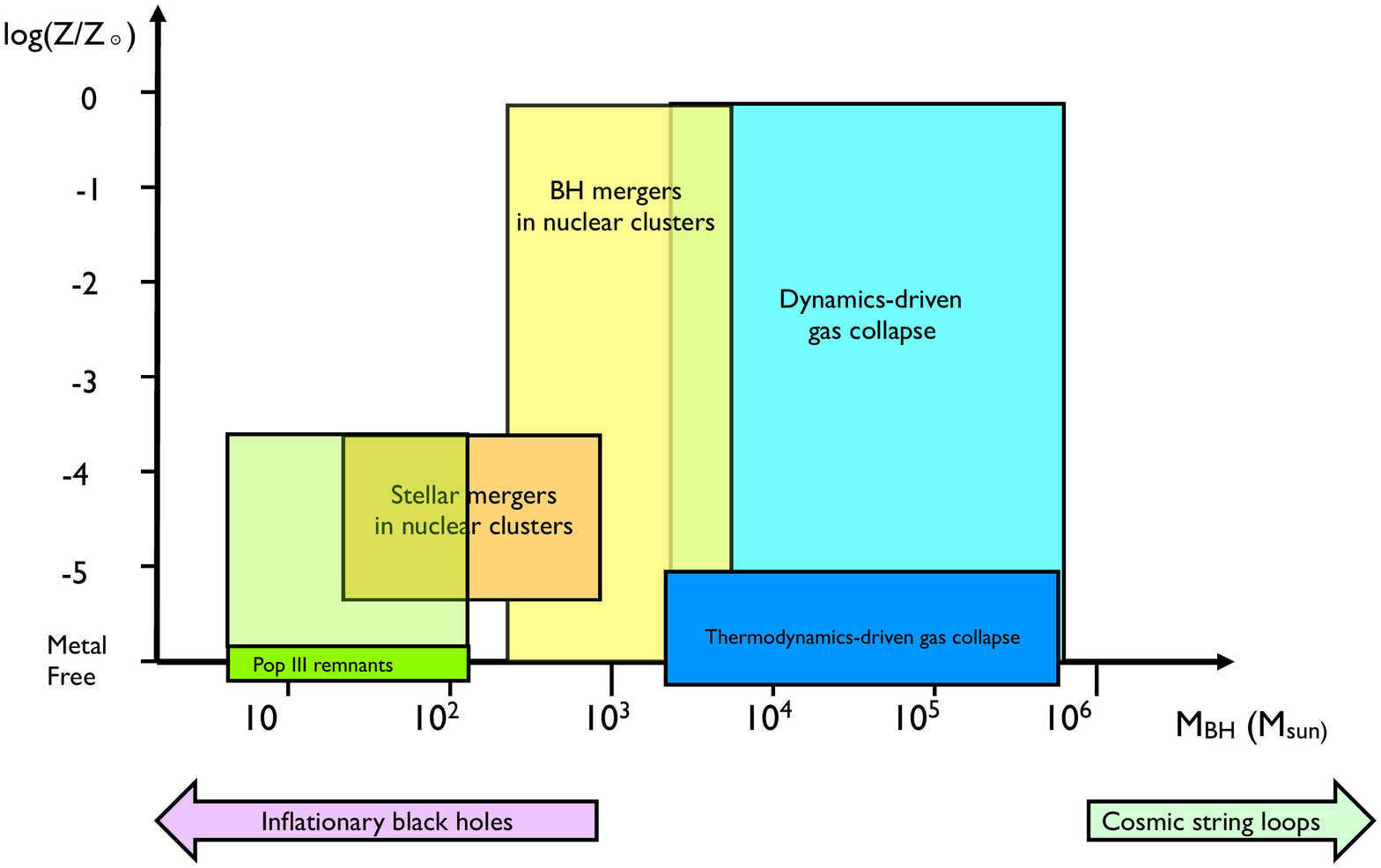} 
 \caption{Cartoon summary of MBH formation models. }
   \label{fig1}
\end{center}
\end{figure}

Different MBH formation mechanisms have been proposed in the literature (see Fig.~\ref{fig1}), starting from \cite{Rees1984}.  Some proposed mechanisms, such as primordial (inflationary) MBHs \citep{Khlopov} and cosmic string loops \citep{2015JCAP...06..007B} are not directly linked to galaxy formation, and I refer the reader to the original papers for details. Recent reviews on the other processes include \cite{Volonteri2010AARV,2013ASSL..396..293H}. 

We will discuss here only one parameter relevant to MBH formation: the metallicity of the gas or stars that go into MBH formation. The crucial reason is related to stellar evolution, and specifically mass losses through winds. \cite{Yungelson2008} study the fate of solar composition stars in the mass range 60-1000 $\msun$. They find that they shed most of their mass via winds and are expected to end their lives as MBHs less massive than $\sim 150 \msun$. At low metallicity, instead, mass loss due to winds is much more reduced, thus increasing the mass of the final remnant \citep{2015MNRAS.451.4086S}. MBH formation mechanisms that require an intermediate phase of (super)massive star, therefore, need to occur in metal poor conditions, unless the star can collapse directly through general relativistic instability \citep{1966ApJ...144..180F,Baumgarte1999,2002ApJ...572L..39S}. Such metallicity requirements do not apply to models that invoke stellar mass MBH mergers \citep{2011ApJ...740L..42D,2012ApJ...755...81M,Lup2014}. \nocite{2015arXiv150404042A}

One important point to keep in mind is that MBH formation must be common enough to account for MBHs in galaxies such as the Milky Way, as well in dwarf galaxies \citep{2012NatCo...3E1304G,2013ApJ...775..116R}. The currently fashionable models of thermodynamics-driven direct collapse at zero \citep[e.g.,][]{2013MNRAS.433.1607L,Regan2014B} or very low metallicity \citep{2015arXiv150907034L} fostered by a substantial Lyman-Werner flux seem to require very stringent conditions, leading to a very low expected number density \citep[and references therein, but see Agarwal et al. 2015]{2015arXiv150705971H}. While this mechanism, predicting very high seed masses, around $10^5 \msun$ \citep{LN2007,2014MNRAS.443.2410F}, is favoured by  timescale arguments for growing high-z quasars, it seems unable to seed run-of-the-mill galaxies. One or more MBH formation mechanisms can occur in the Universe, though, they are not necessarily mutually exclusive \citep[e.g.,][]{VB2010,Devecchi2012}.

\section{BH feeding}

\begin{figure}[t]
\begin{center}
 \includegraphics[width=\columnwidth]{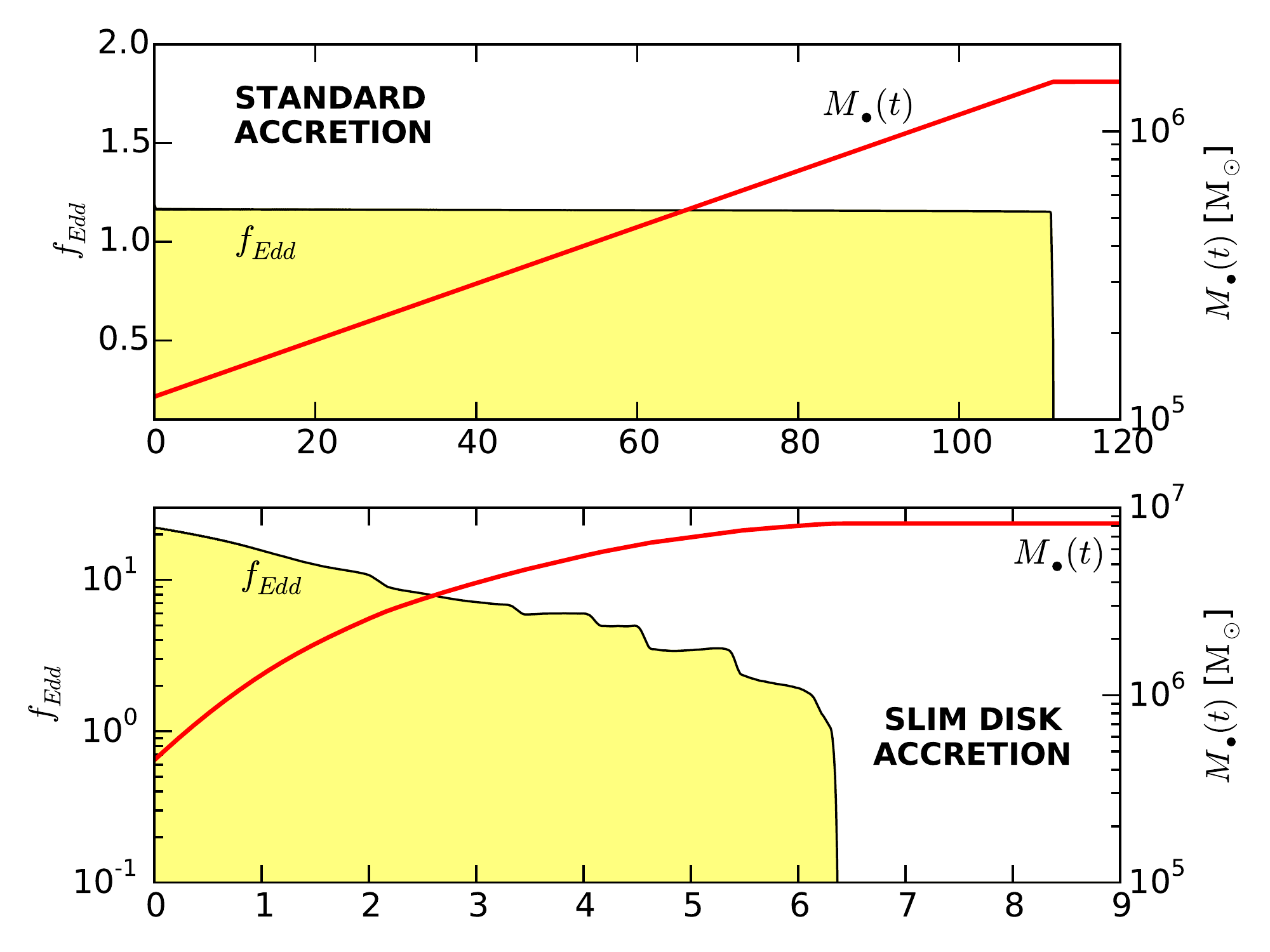} 
 \caption{Time evolution of the Eddington ratio $f_{\rm Edd}$ and of the black hole mass $M_{\bullet}$, in the standard case, where  $L = f_{\rm Edd} L_{\rm Edd}$ (top) and the slim disc case, where $L \sim \ln(f_{\rm Edd}) L_{\rm Edd}$ (bottom), so that luminosity remains sub-Eddington, while accretion is super-critical. }
   \label{fig2}
\end{center}
\end{figure}

The very bright quasars with $z>6$ require to be fed at the Eddington level for their whole lifetime. Are such long-lived accretion events possible? \cite{Lietal2006}, using  series of successive mergers extracted from a cosmological run,  suggested that merger-driven accretion does the trick. \cite{2012ApJ...745L..29D} propose instead direct accretion from the cosmic cold flows. They use direct cosmological simulations, but with relatively low resolution ($\sim$1~kpc). At the other extreme, \cite{Bournaud2011}, use very high resolution (1~pc) but isolated disc galaxy simulations to suggest that accretion is driven by disc instabilities in high-z galaxies. Finally, \cite{duboisetal12angmom} use cosmological zooms, with a maximum resolution of $\sim 10$~pc to find that direct accretion of cosmic gas dominates early on, disc feeding takes over at later times, and galaxy mergers become important even later on. 

In all these studies, the simulation provides information on the gas inflowing from the galaxy. The accretion rate on the MBH is calculated from this inflow rate, with an explicit cap at the Eddington luminosity. In a real galaxy, how does it know that it has to feed the MBH {\it exactly} at the rate resulting in the Eddington luminosity? It is important to stress that super-critical accretion does not necessarily imply highly super-Eddington luminosity. At high accretion rates the thermodynamics and geometry of the accretion disc changes \citep[e.g., slim discs,][]{Abramowicz1988}. The material is so dense and optically thick that  photons are advected inward with the gas, rather than diffuse out. Radiation is trapped, and luminosity highly suppressed \citep{Begelman1979}: $L \propto \ln \dot{M}$ instead of $L \propto \dot{M}$. Short periods of super-critical accretion are sufficient to ease the constraints on the growth of billion solar masses MBHs at $z>6$ \citep[e.g.,][]{Volonteri2005,2015ApJ...804..148V}. As proof of concept, \cite{2015MNRAS.452.1922P} perform 1D simulation including radiation transfer of a high-z halo feeding a seed MBH. The radiation-related quantities are integrated over frequencies with matter and radiation coupled via Thomson (electron) scattering and bound-free interactions. The accretion rate is calculated at the innermost cell (0.002~pc) and the accretion and radiative efficiencies are implemented in two ways. In one series of runs, the ``standard" case luminosity is calculated with a fixed matter-energy conversion factor ($\epsilon = 0.1$), and $L = f_{\rm Edd} L_{\rm Edd}$. In the ``slim disc" case, only a fraction of the emitted luminosity escapes to infinity, and the disc becomes radiatively inefficient, with $\epsilon \sim 0.04$ \citep{Mineshige2000} and  $L \sim \ln(f_{\rm Edd}) L_{\rm Edd}$. The effective accretion on the MBH is modulated by gas inflows (feeding the MBH) and radiation pressure (sweeping back the gas). The evolution of MBH masses and the Eddington rate are shown in Fig.~\ref{fig2}. While in standard  case the  MBH accretes $\sim 5\%-15 \%$ of the available gas in $\sim 100 \, \mathrm{Myr}$, with outflows modulating the gas inflows, in the slim disc case the MBH accretes $> 80\%$ of the available gas in $\sim 10 \, \mathrm{Myr}$, with outflows playing a negligible role.

\section{BH feedback}

\begin{figure}[t]
  \centering
  \begin{minipage}{\textwidth}
    \includegraphics[width=\textwidth]{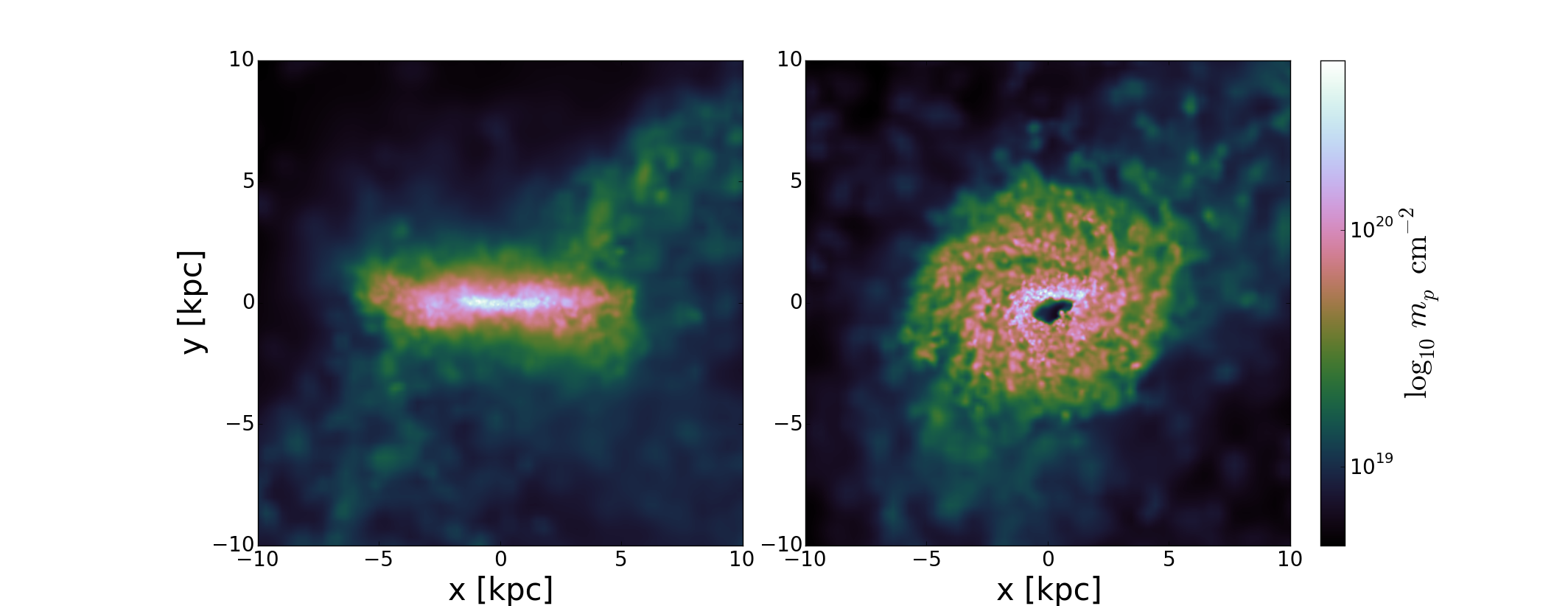}
  \end{minipage}
  \vfill
  \begin{minipage}{\textwidth}
    \includegraphics[width=\textwidth]{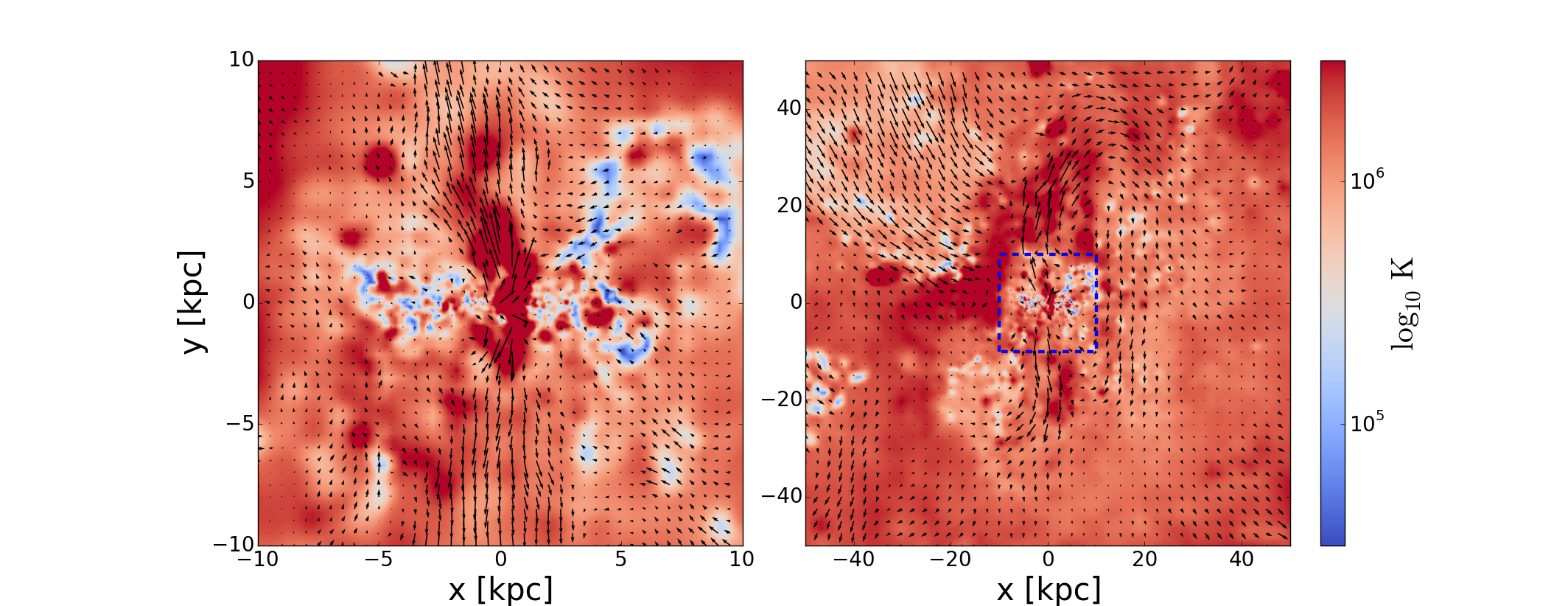}
    \caption{AGN feedback does not destroy galaxies. They drive large-scale outflows that regulate and eventually stop the flow of cold gas onto their host galaxy.}
  \end{minipage}
   \label{fig3}
\end{figure}

Feedback from  MBHs is an important ingredient to understanding the relationship between galaxy and black hole growth \citep{1998A&A...331L...1S}. Winds from active MBHs find the easiest path through low-density gas, avoiding the dense parts of the galaxy disc  \citep{2014MNRAS.441.1615G,2015ApJ...800...19R}. Cosmological simulations show that, rather than tearing apart the galaxy, the AGN create large-scale polar outflows that eventually destroy the cold circumgalactic medium, leaving a hot halo and strangling the host galaxy's supply of cold gas \citep{Dubois2013,2014MNRAS.444.2355C,Tremmel}. This is supported by observations, as star forming galaxies are common hosts of bright AGN \citep{2013A&A...560A..72R,Mullaney2012}. AGN feedback is able to explain the inefficient star formation observed in massive galaxies, causing the turn-over in stellar mass-halo mass relationships  \citep[e.g.,][]{Croton2006,Tremmel}.

MBHs also affect their own gas supply through feedback, possibly stunting their growth at early times \citep{Alvarez2009,Milos2009,2011MNRAS.410..919J,2012ApJ...747....9P,2014ApJ...797..139A} and eventually reaching a state of self-regulation at later times. The onset and nature of this self-regulation phase is, at least in part, governed by SN feedback and nearby gas dynamics  \citep{2015MNRAS.452.1502D,Habouzit,Tremmel}.

AGN can also have positive feedback on their surroundings  \citep[e.g.,][]{2013ApJ...774...66Z,2015A&A...582A..63C}, triggering star formation, for instance because of shock pressure-enhanced star formation \citep{2013ApJ...763L..18W,Silk2013,2015arXiv150700730B,2012MNRAS.425..438G}.

An important caveat is that the implementation of AGN feedback is still very heuristic. Simulations including radiative transfer \citep{Bieri} and realistic jets \citep{Cielo}, as well as all the physics of galaxy formation, will be paramount to have an accurate understanding of how feedback affects the MBH itself and the galaxy.

\section{BHs and galaxies}
\begin{figure}[t]
\begin{center}
 \includegraphics[width=\columnwidth]{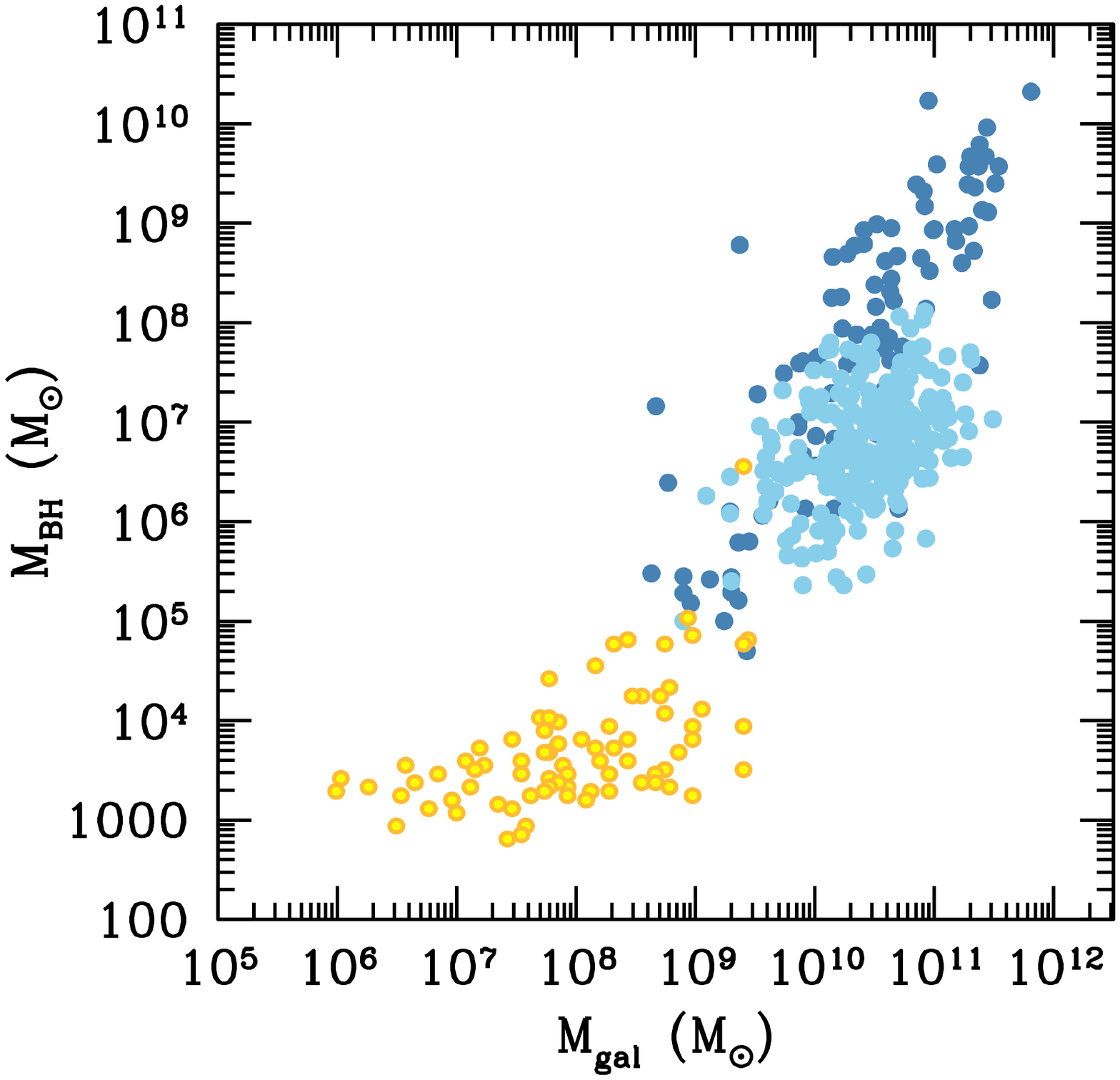} 
 \caption{BH mass versus galaxy stellar mass. Dark and light blue points are quiescent MBHs  and low-luminosity AGN at $z=0$ from \cite{2015arXiv150806274R}. Yellow points are MBHs in a cosmological simulation (using the adaptive mesh refinement code RAMSES, with a maximum resolution of 80~pc, Habouzit et al. in prep). MBHs in small galaxies are unable to grow because of the interplay between SN and AGN feedback, very effective in small galaxies' potential wells. Once a galaxy has become sufficiently massive to contrast outflows, its MBH can thrive.}
   \label{fig4}
\end{center}
\end{figure}

The constraints on the relationship between MBH masses and galaxies properties at $z\simeq6$ are few, and seem to provide conflicting results.  (i) There seems to be little correlation between MBH mass and host properties, or at least a much larger scatter \citep{Wang2010,2015ApJ...801..123W}, (ii) typically MBHs are `over-massive' at fixed galaxy mass/velocity dispersion compared to their $z=0$ counterparts \citep[e.g., Walter et al. 2004; at lower redshift see also][]{Mclure2004,Shields2006, Decarli2010,Merloni2010}, but (iii) analysis of the MBH mass/luminosity function and clustering suggests that either many massive galaxies do not have MBHs, or these MBHs are less massive than expected \citep{Willott2010,2015MNRAS.448.3167W}.  As a result of point (ii), most authors propose that there is {\it positive} evolution in the MBH mass-galaxy relationships, and quantify it as a change in {\it normalization}, in the sense that at fixed galaxy properties (e.g. velocity dispersion, stellar mass), MBHs at high redshift are more massive than today. However, this is inconsistent with point (iii) above. \cite{2011MNRAS.417.2085V} propose that if the evolution of the correlations is in {\it slope} rather than {\it normalization}, with a tilt such that MBHs in small galaxies are under-massive and MBHs in large galaxies are over-massive, then one can reconcile the observational results (i)-(ii) and (iii) above. 

While observations are still limited, cosmological simulations seem indeed to find that MBHs in small galaxies  are unable to grow and remain ``stuck" at low mass, as shown in Fig.~\ref{fig4}, while MBHs in the most massive galaxies can grow beyond the correlations between MBH mass and host properties found at $z=0$ \citep{2015MNRAS.452..575S,2015MNRAS.454..913D,Habouzit}. Indeed  low-mass galaxies are a fragile environment, and supernova feedback is sufficient to energize the gas and suppress MBH accretion \citep{2015MNRAS.452.1502D}. MBHs are also easily perturbed from the center of small galaxies and experience long orbital decay timescales, making it more difficult for them to grow efficiently \citep{2015MNRAS.451.1868T}.

\section{Conclusions}
High redshift MBHs and quasars represent both a theoretical and observational challenge. The current sample is limited to the brightest sources, powered by extremely massive MBHs. One should appreciate that theory and observations have the same limitation: either one has a large, shallow survey (large, low-resolution simulation), or a narrow, deep survey (small, high-resolution simulation). The number density of these quasars is so low, $\sim$ one per Gpc$^3$, that comparable volumes must be simulated in order to catch a glimpse of their evolution. Such large volumes are limited in terms of physical resolution, but we are currently reaching a good balance between size and spatial information \citep{2015arXiv150406619F}.  Several telescopes in the near future promise to greatly increase the sensitivity, thus being able to detect lower luminosity AGN, hopefully powered by lower mass MBHs \citep{2015arXiv150605299P}. {\it JWST}, {\it ATHENA} and {\it SKA} are all geared towards finding and studying the high redshift Universe, and the MBHs and galaxies that inhabit it. Meanwhile, {\it ALMA} can provide us with unprecedented information on the dynamics and kinematics of molecular gas in the galaxies hosting quasars and AGN. On the theoretical side, large volumes at high-resolution are becoming a reality, and we can now shift the attention from reproducing statistical results, to try and understand the {physics} of MBH formation, feeding and feedback. 

\section*{Acknowledgments}
MV  acknowledges funding from the European Research Council under the European 
Community's Seventh Framework Programme (FP7/2007-2013 Grant Agreement no.\ 614199, project ``BLACK'').  

\bibliographystyle{mn2e}

\end{document}